# An accelerating approach of designing ferromagnetic materials via machine learning modeling of magnetic ground state and Curie temperature

**T. Long, N. M. Fortunato, Yixuan Zhang, O. Gutfleisch, and H. Zhang***

Institute of Materials Science, Technical University of Darmstadt, Darmstadt 64287, Germany

*Corresponding to Hongbin Zhang, hzhang@tmm.tu-darmstadt.de

## Abstract

Magnetic materials have a plethora of applications ranging from informatics to energy harvesting and conversion. However, such functionalities are limited by the magnetic ordering temperature. In this work, we performed machine learning on the magnetic ground state and the Curie temperature ($T_C$), using generic chemical and crystal structural descriptors. Based on a database of 2805 known intermetallic compounds, a random forest model is trained to classify ferromagnetic and antiferromagnetic compounds and to do regression on the $T_C$ for the ferromagnets. The resulting accuracy is about 86% for classification and 92% for regression (with a mean absolute error of 58K). Composition based features are sufficient for both classification and regression, whereas structural descriptors improve the performance. Finally, we predict the magnetic ordering and $T_C$ for all the intermetallic magnetic materials in the Materials Project. Our work paves the way to accelerate the development of magnetic materials for technological applications.

## Introduction

The continued growth of the global population has raised issues about sustainability and energy future, demanding improved efficiency of electricity production and consumption. Magnetic materials have a wide spectrum of applications, particularly in efficient energy harvesting, conversion, and utilization.[1,2] Specifically, permanent magnets (PMs) are the key components for the energy related technologies, such as conventional generators, e-mobility, automatization and refrigeration.[3,4] Currently, two classes of PMs, namely, the ferrites and AlNiCo, and the high performance PMs based on Nd-Fe-B and Sm-Co are widely used, with a gap in between to be filled by novel PMs, ideally those without critical elements such as heavy rare earths. Moreover, FM materials have been widely applied in spintronics, such as sensing, memory and logic, whereas the emerging antiferromagnetic (AFM) spintronics have recently drawn intense attention.[4,5] Two fundamental properties desired for promising candidate magnetic materials are a ferromagnetic (FM) ground state with strong magnetization and a high Curie temperature ($T_C$) which governs the temperature range of functioning. These properties are also important for magnetic refrigeration which promises enhanced energy efficiency over the conventional cooling technologies.[6]

Although $T_C$ is readily experimentally measurable, synthesis and optimization of real materials are time-consuming and mostly done based on trial and error. Thus, the development of a methodology to accelerate the development of magnetic materials with a theoretical pre-screening is of natural interest. Typical theoretical approaches to evaluate $T_C$ rely on the parameterization of density functional theory (DFT) electronic structure to construct a Heisenberg Hamiltonian, which can be solved via atomistic Monte Carlo simulation. This approach fails even for elemental metal like Co and Ni, due to the strong itinerant nature of magnetism therein.[7,8] Moreover, DFT is not sufficient in describing the strongly correlated 4f electrons in rare-earths,[9] while the orbital dependent functional (e.g., DFT+U) treatment is often chosen to fit to experiments. The state-of-the-art DFT plus dynamical mean field theory (DMFT) method can be applied to tackle the electronic correlation problem but is numerically expensive. Whereas the $T_C$ evaluated for bcc Fe based on DFT+DMFT is 50% off the experimental value.[10,11] Therefore, there is a great impetus for a predictive approach to obtain $T_C$, which is applicable to compounds with arbitrary compositions and crystal structures.

Machine learning is an emerging tool in materials science, being applied successfully to model the thermodynamic stability,[12] band gap,[13] elastic properties,[14] inter-atomic potentials[15] and in predicting potential high temperature superconductors.[16] However, regression models to predict ordering temperature of magnetic materials have only been reported in a limited scope while classification models to distinguish AFM and FM are absent in literature to the best of our knowledge. Sanvito et al. trained a linear regression model over 40 intermetallic Heusler alloys (with experimental $T_C$), and made predictions for another 20 compounds. By validating experiments, they discovered $Co_2MnTi$ with a remarkably high $T_C$ of 900K.[17] Dam et al. focused on selecting the best features for predicting $T_C$ of binary 3d-4f intermetallic compounds by applying Gaussian kernel regression on 108 compounds. The add-one-in test accuracy can reach above 95% when only eight descriptors are used, with the rare-earth concentration being the most relevant.[18]

In this work, we develop a FM/AFM classification model along with a regression model to predict the $T_C$ for intermetallic FM compounds, using the random forest (RF) method. These models are then used to identify the magnetic ground state of 5183 magnetic intermetallic compounds from the Materials Project database and to predict the $T_C$ of those classified as FM. It is demonstrated that our machine learning framework is efficient and predictive, and can be used to accelerate the screening for FM compounds which are promising for spintronics and permanent magnets applications.

# Results

## Data

Using the AtomWork database,[19] 1749 FM and 1056 AFM inter-metallic compounds are collected, where oxides, sulfites, chlorides, and fluorides having been excluded, along with compounds without either of Cr, Mn, Fe, Co, and Ni atoms, which are the typical magnetic atoms in transition metal based intermetallic magnetic materials. The corresponding crystal structures are collected from AtomWork and Inorganic Crystal Structure Database (ICSD).[20] For compounds with multiple magnetic phase transitions, the critical temperature is defined as the magnetic transition temperature from a disordered paramagnetic state to an ordered FM state. In this way, materials with the first-order magneto-volume, magneto-structural, and temperature dependent spin-reorientation transitions are excluded for the current work.

The distribution of experimental FM ordering temperatures is shown in Fig. 1(A). It is dominated by compounds with low $T_C$, with a maximum value around 1400 K, e.g., 1410 K and 1388 K for Co in face-centered-cubic (space group 225) and hexagonal-closed-pack (space group 194) structures, respectively. The element resolved $T_C$ distributions are highlighted in Fig. 1(B-D) for Fe-, Co-, and Mn-based compounds, whereas the $T_C$ for Ni-based compounds are mostly in the low-temperature range (Fig. S1(A)) and there are limited number of Cr-based ferromagnetic materials (Fig. S1(B)). It is clear that all compounds with $T_C$ higher than 1200K are Co-based (Fig. 1(D)), the Fe-based compounds (Fig. 1(C)) consist of 1/3 of the database with $T_C$ normally distributed around 600K, while the $T_C$ of Mn-based compounds (Fig. 1(B)) are mostly found at relative low temperature (with a peak at 300 K) range. Thus, the Fe- and Co-based compounds are optimal for high temperature applications, and in the room temperature range all three classes are interesting.

Moreover, the distribution of experimental AFM ordering Neel temperatures ($T_N$) is shown in Fig. S2(A), where $T_N$ of most compounds are less than 100K. The element resolved $T_N$ distributions in Fig. S2(B-F) indicate that Fe- and Mn- based compounds are more suitable for high temperature application. It is noted that there are many more AFM compounds such as oxides, thus the collection of AFM intermetallic compounds in this work serves only for classification and we save the regression of $T_N$ for future study. We have also collected 5193 magnetic intermetallic compounds from Materials Project,[21] in order to make predictions by applying the machine learning models for magnetic ground state classification and $T_C$ regression, as discussed in detail below.

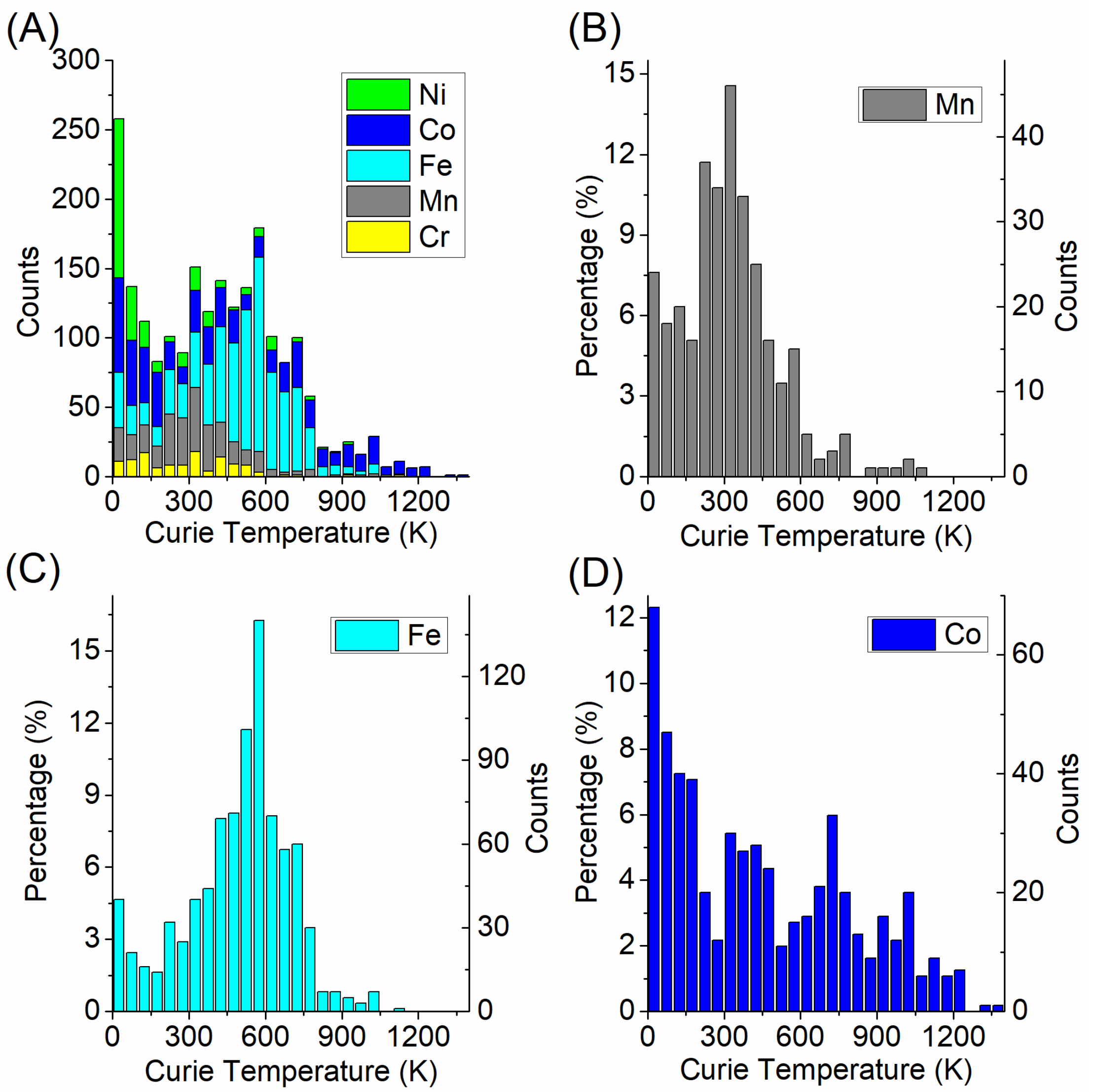

**Fig. 1. Distribution of $T_C$ for the FM database. (A)** Histogram of $T_C$ for 1749 FM materials in the database. Green, blue, cyan, gray and yellow represents Ni, Co, Fe, Mn and Cr based compounds, respectively. **(B)** Histogram of Fe based FM compound categorized by $T_C$. **(C)** Histogram of Co based FM compound categorized by $T_C$. **(D)** Histogram of Mn based FM compound categorized by $T_C$. The bin size is fixed to be 50K.

## Descriptors

The Materials Agnostic Platform for Informatics and Exploration (MAGPIE)[22] proposed by Ward et al. is used to obtain the chemical descriptors, including 4 categories: stoichiometric attributes, element properties statistic, electron structure attributes and ionic compound attributes. In this work, we regroup them in the following 5 classes, namely, norm ($L^p$ norms of the fractions), magnetic moment of the constituting elements, atomic number, valence electrons and other chemical descriptors. (Please check details in Methods and Table S1 and S2) These descriptors are collectively labeled as CHEM. As for structural descriptors (labeled as STR), Smooth Overlap of Atomic Positions (SOAP) is used to describe the local crystalline environment such as coordination and distance between atoms. [23] Space group number is considered as a structural descriptor as well. In total, 139 (25) CHEM (STR) descriptors for each compound are used. In order to get a better understanding of the relative feature importance and the underlying physical picture, we considered two models with variations of the included features labeled as CHEM+STR and CHEM.

## Classification

While the prediction of whether a material is magnetic or not can be straightforwardly done in DFT, the question of being AFM or FM is more complex. For instance, the AFM ground state is mostly set aside in the Materials Project, despite the thermodynamic stability and electronic properties hinge on the magnetic configurations. This is because that the number of possible AFM states, especially after considering non-collinear magnetic configurations,[24] would make high-throughput calculations intractable.[25]

To enable the prediction of the magnetic ordering in a computationally inexpensive fashion, we perform a classification of the AFM or FM ground state. The training set consisted of 90% of the database and 10% are used for validation. The quality of the classification model is judged in terms of the following statistic metrics.[26] By looking at the positives (tp), false positives (fp), false negatives (fn) and true negatives (tn), the accuracy, precision, recall and F1 can be evaluated as following:

$$accuracy = \frac{tp + tn}{tp + tn + fp + fn}$$

$$precision = \frac{tp}{tp + fp}$$

$$recall = \frac{tp}{tp + fn}$$

$$F_1 = 2 \times \frac{precision \times recall}{precision + recall}$$

with true being FM and false AFM. The accuracy represents the overall quality of the prediction, the precision is the proportion of those correctly classified as FM within all classified as FM, the recall is the proportion of those correctly identified as FM with all known to be FM, and the F1 bridges the recall and precision metrics, denoting if there is a bias towards classifying one label. The confusion matrix (CM) is a table that represents the instances in a predicted class versus the ones in the actual class, as shown in Fig. 2(A), together with the resulting metrics for 10 cross-validation sets plotted in Fig. 2(B).

Obviously, the best accuracy for classification is 86%, that is, 88.8% FM and 82.4% AFM compounds are correctly classified. It is achieved by taking all chemical and structural features as descriptors, i.e., the CHEM+STR model. This combined with an F1 score of 89% (Fig. 2(B)) indicates good predictability, with a slight bias towards predicting FM which might be due to the unbalanced number of FM/AFM compounds in the database. By performing 10-fold cross validation (Fig. 2(B)), the average accuracy is about 82%, meaning that the RF model has neither overfitting nor biased sampling.

Interestingly, in the CHEM+STR model, the descriptor group valence electrons is selected as the most important feature, contributing 47% of feature importance, while the second most important descriptor group, SOAP, only constitutes 15%, followed by the atomic number and magnetic moment of the constituting elements. The contribution of valence electrons demonstrates that intrinsic properties of element have the strongest significance in predicting the magnetic ordering. To confirm this, the CHEM model is considered, which eliminates an explicit description of the local crystalline environment, leading to an accuracy of 82% and an F1 score of 86% (Table S3) where the order of importance for the chemical features remains the same, reinforcing their importance relative to each other. Excluding the structural descriptor only results in a 4% drop in accuracy, indicating that the local environment does not affect the magnetic ordering directly, but exerts influence on the valence electrons of atoms and disturbs the magnetic ordering. However, the necessity of considering interaction between atoms in magnetic ordering cannot be overlooked.

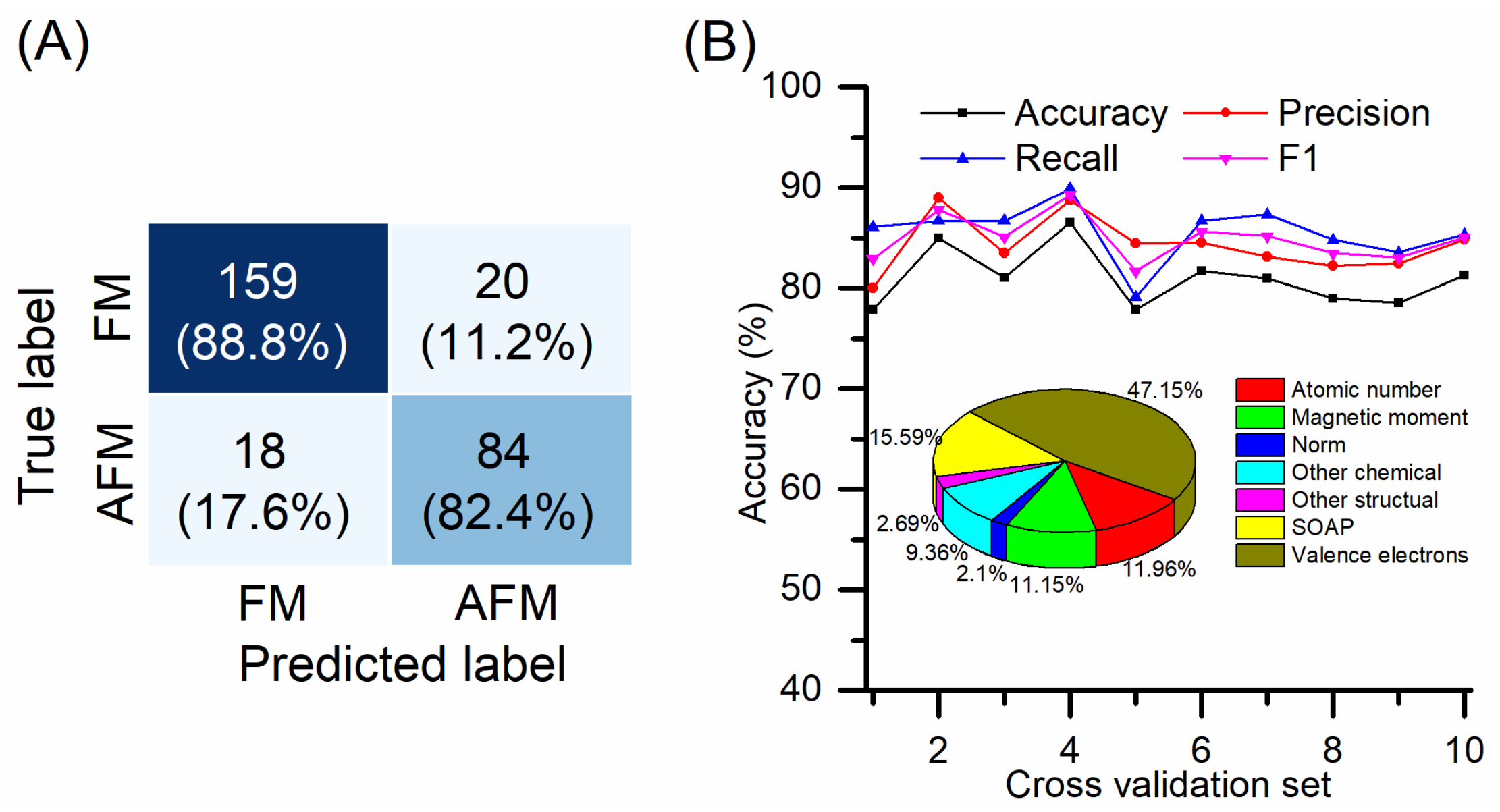

**Fig. 2. Performance of the classification model. (A)**: Confusion matrix of FM and AFM classification test set of 281 compounds. **(B)**: Statistical metrics for 10-folder cross validation.

## Regression

Turning now to the regression of $T_C$, which is done using the RF with 90%/10% partition for training/validation of the 1749 FM compounds, the best $R^2$ obtained from the validation is as high as 92% using the CHEM+STR descriptors, indicating very good agreement between the experimental and predicted values. As shown in Fig. 3(A), the corresponding mean absolute error (MAE) is 58 K. The agreement has to be weighed against the variation of $T_C$ in experiments due to differences in composition, synthesis and measurement techniques, which contributes to the error. It is noted that, compared with those obtained based on DFT calculations, the machine learned values are more accurate. [7–9]

The valence electron features are still assigned with the highest importance of 41%, while that for the magnetic moment of the constituting elements increases to 25% corresponding to the linear relationship between magnetic moment per atom with the total magnetization. Interestingly, the importance of SOAP drops to only 9%, indicating that the local crystalline environment has less importance when FM ordering is determined. When compared with the CHEM descriptors, the effect of including the crystal structure is again marginal but noticeable. For instance, using only the CHEM descriptors results in a $R^2$ of 90% and an MAE of 60K (Table S4). Nevertheless, the CHEM only model must fail for edge cases, where different phases/volumes of the same compound have different magnetic orderings or $T_C$. For instance, $Fe_3Nb$ with space group number 225 has a predicted $T_C$ of 939K while the predicted $T_C$ is only 373.67K with space group number 139 using CHEM+STR descriptors. However, Heusler alloys with magneto-structural transitions like $Ni_2MnGa$ exhibit the same predicted $T_C$ in both cubic and tetragonal structures, probably due to the lack of data in training, e.g., only 32 compounds in the database have isomers.

Furthermore, in order to test the convergence with respect to the feature space, we performed active learning (AL), starting with 10% of the training data and take automatically 10% more samples data which are the most outliers in the rest training set. In this way, the feature space that has not been covered by the previous training set will be included, enabling to obtain higher accuracy with less data.[27] As shown in Fig. S3, using only 50% of the data, the resulting accuracy based on AL is already comparable with the standard model. Such training accelerating is even significant when approximately 10% data selected by AL achieves an $R^2$ of 84% compared with random selection only at 72%. Therefore, covering larger region of the feature space will make the regression modeling more robust. In addition, our database is not guaranteed to be completed, the AL algorithm will make it possible to take further experimental cases into consideration.

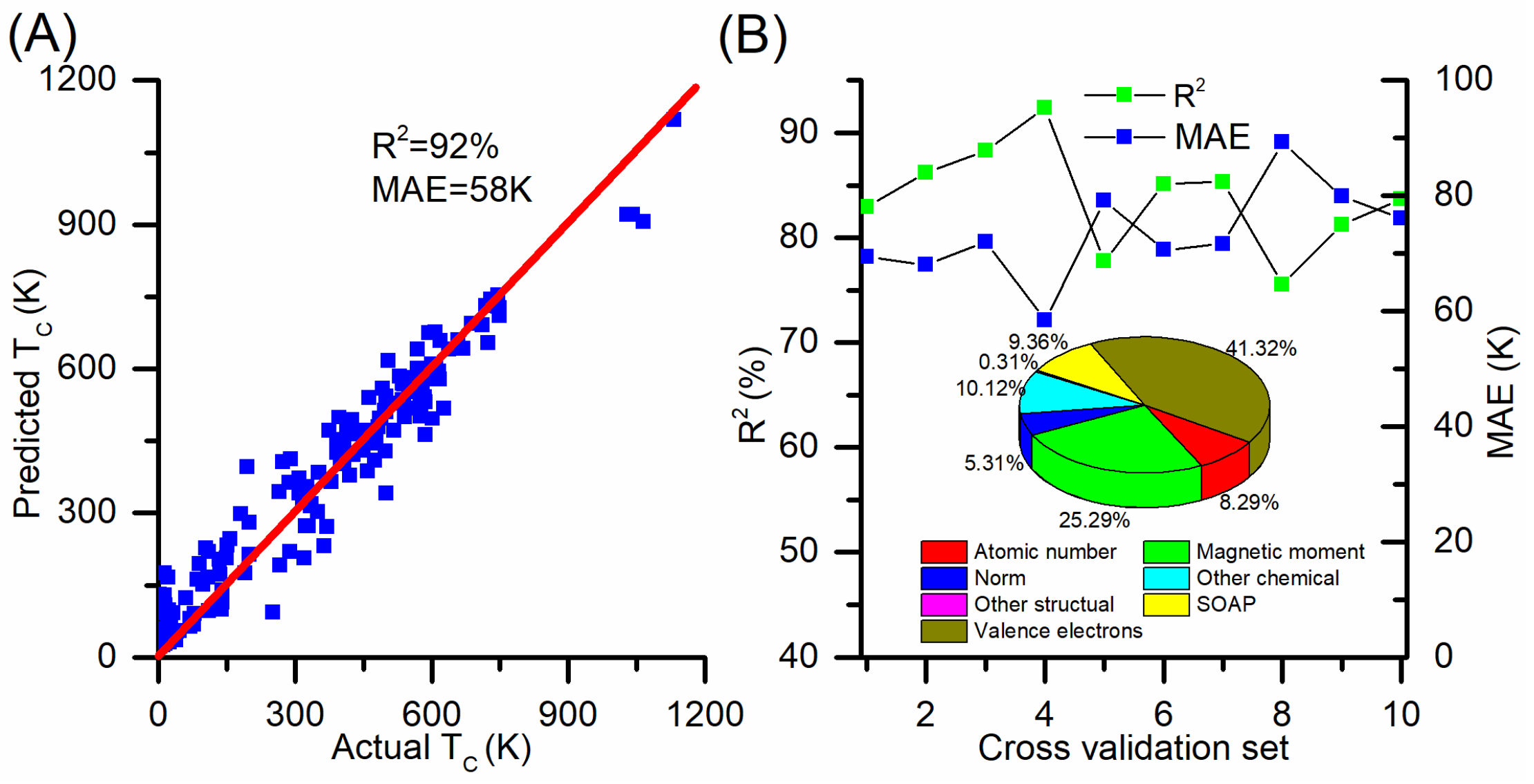

**Fig. 3. Regression model performance and feature importance. (A)** Predicted vs. experiment $T_C$ for test set in general regression model. **(B)** 10-folder cross validation and pie chart of feature importance.

## Prediction

Using the model based on CHEM+STR, we performed AFM/FM classification of the 5193 intermetallic compounds from Materials Project, leading to 2884 (2309) FM (AFM) (see Data file S1 in the Supplementary). Fig. 4 shows the predicted $T_C$ for the classified FM compounds, ranging up to 1280 K. Obviously, the compounds with $T_C$ higher than 600 K are again mostly Co and Fe based, and the element-wise distribution is comparable to that of our training set as shown in Fig. S4. The compounds with predict $T_C$ > 1100 K are listed in Table 1, which are all Co-based. It is found that the predicted $T_C$ is in good agreement with the values from the literature, which have been collected after obtaining the prediction. For instance, the largest difference is around 130K for $Co_{17}Tb_2$, where the predicted value is about 1283.7K while the experiment value is 1150K [28]; whereas the smallest difference is about 10 K for $Co_{17}PrYb$ with experimental (predicted) $T_C$ being 1178K[29] (1167.2 K). We note the total computational time for the classification and regression of these compounds is a few seconds, making it trivially inexpensive compared with the computational effort that would be required by DFT.

**Table 1. List of potential high $T_C$ (>1100K) FM compounds by this work.**

| MP-id | Formula | Predicted $T_C$ (K) | Real $T_C$ (K) | Space Group No. |
|---|---|---|---|---|
| mp-1072089 | Co | Around 1283.67 | - | 227 |
| mp-669382 | Co | Around 1283.67 | - | 186 |
| mp-1193227 | Co | Around 1283.67 | - | 136 |
| mp-1096987 | $Co_9Fe$ | Around 1283.67 | - | 123 |
| mp-1201816 | $Co_{17}Gd_2$ | Around 1283.67 | 1200[30] | 194 |
| mp-1204082 | $Co_{17}Lu_2$ | Around 1283.67 | 1192[31] | 194 |
| mp-1195194 | $Co_{17}Np_2$ | Around 1283.67 | - | 194 |
| mp-568820 | $Co_{17}Pu_2$ | Around 1283.67 | - | 194 |
| mp-1094061 | $Co_{12}Sm$ | Around 1283.67 | - | 139 |
| mp-16932 | $Co_{17}Th_2$ | Around 1283.67 | - | 166 |
| mp-1199370 | $Co_{17}Tb_2$ | Around 1283.67 | 1150[28] | 194 |
| mp-1196360 | $Co_{17}Tm_2$ | Around 1283.67 | 1170[28] | 194 |
| mp-1219785 | $Co_{17}PrSm$ | Around 1167.17 | 1200[32] | 160 |
| mp-1219295 | $Co_{17}GdSm$ | Around 1167.17 | 1200[32] | 160 |
| mp-1220026 | $Co_{17}ErPr$ | Around 1167.17 | 1167[29] | 160 |
| mp-1200096 | $Co_{17}Sm_2$ | Around 1167.17 | 1150[28] | 194 |
| mp-1215870 | $Co_{17}PrYb$ | Around 1167.17 | 1178[29] | 194 |
| mp-1199900 | $Co_{17}Yb_2$ | Around 1167.17 | 1176[29] | 194 |
| mp-1216133 | $Co_{33}Yb_4Zr$ | Around 1167.17 | 1175[33] | 156 |
| mp-1215883 | $Co_{34}Pr_3Yb$ | Around 1167.17 | 1173[29] | 8 |
| mp-1219047 | $Co_{17}SmY$ | Around 1167.17 | 1200[32] | 160 |
| mp-356 | $Co_{17}Nd_2$ | Around 1167.17 | 1125[34] | 166 |
| mp-1224958 | $Co_{33}La_4Ta$ | Around 1167.17 | - | 8 |
| mp-1226612 | $CeCo_{17}Y$ | Around 1167.17 | - | 160 |
| mp-1105621 | $Co_{17}Pr_2$ | Around 1167.17 | - | 166 |
| mp-1220026 | $Co_{17}ErPr$ | Around 1167.17 | - | 160 |

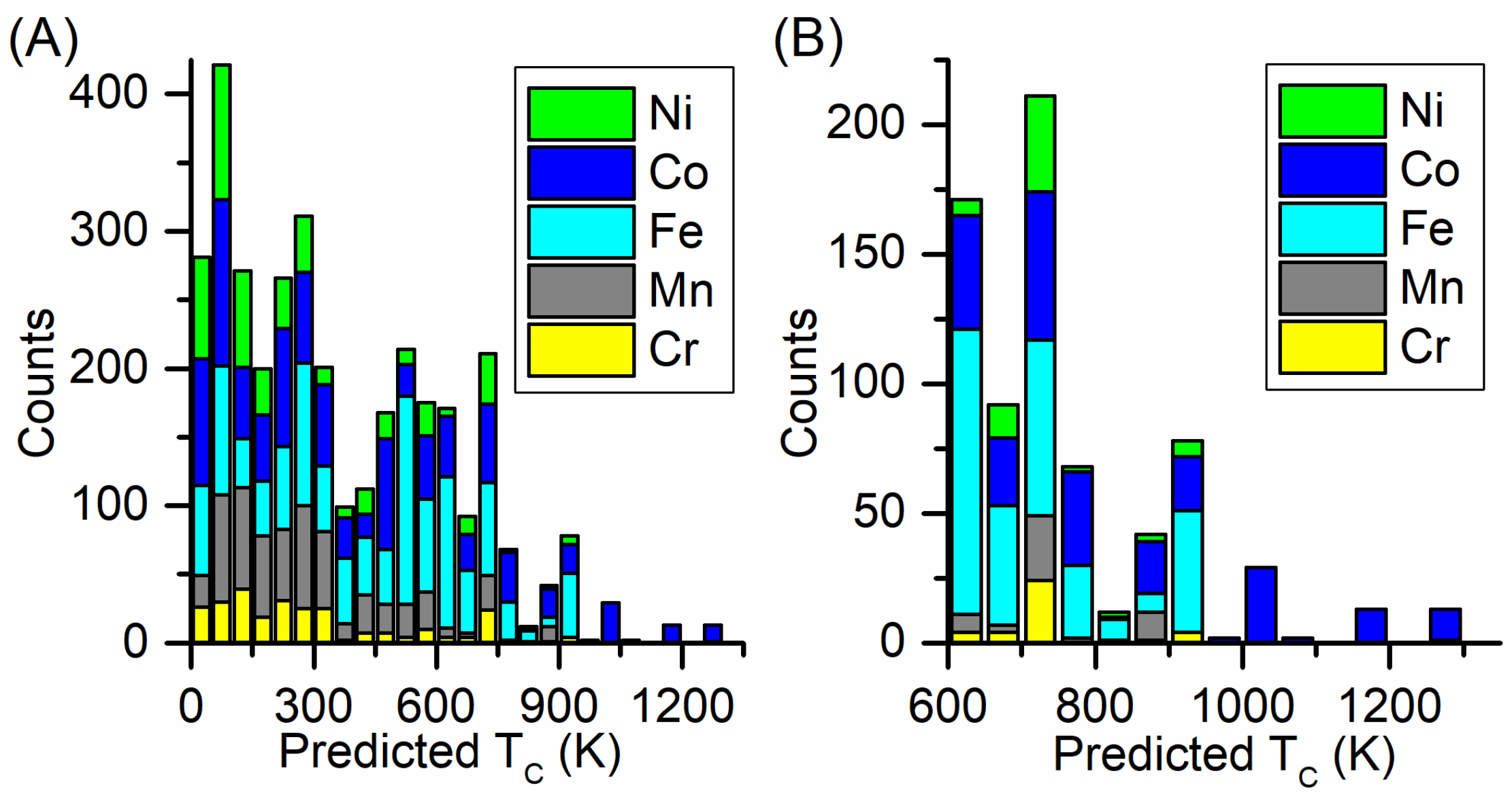

**Fig. 4. Histogram of $T_C$ prediction. (A)** Histogram of predicted $T_C$ for 2884 FM materials in the database. Green, blue, cyan, gray and yellow represents Ni, Co, Fe, Mn and Cr based compounds, respectively. **(B)** The same as (A) but with predicted $T_C$ higher than 600K.

## Discussion

It is demonstrated that machine learning using the RF algorithm is able to distinguish materials with FM and AFM ordering, and further predict the $T_C$ of FM compounds. This solves two critical problems in designing magnetic materials. For classification, the accuracy reaches 86% (82%) using chemistry plus structure (chemistry only) as descriptors. This outperforms the DFT calculations[25], which are applied on a selected set of compounds. For the resulting FM compounds, the magnetization can be straightforwardly evaluated using DFT. Furthermore, the $T_C$ can be accurately modeled with $R^2 \cong 92\%$ and MAE about 58K. This enables us to reduce the number of candidates for further characterization. For instance, the magnetocrystalline anisotropy can be evaluated in a high throughput way,[35] which sets an upper limit for the coercivity. Thus, the machine learning model developed in this work in conjunction with DFT enables us to get all three essential intrinsic magnetic properties evaluated.

Since machine learning is able to capture the mechanism behind magnetic ordering from the statistical point of view, one interesting question is to apply the same modeling on AFM compounds to predict the Neel temperature. As the descriptors we used are robust, as suggested by comparable accuracy with different sets of descriptors, we suspect that our methods are applicable to predict the Neel temperature of AFM compounds as well. However, the AFM magnetic ground states are not uniquely defined, which requires additional development based on either machine learning modeling or high throughput DFT calculations.

In conclusion, we have developed a robust machine learning framework which allows screening of the magnetic ground state and Curie temperature of FM compounds. This paves the way to develop FM materials with systematic characterization of the intrinsic magnetic properties, with the help of further high throughput DFT calculations.

## Acknowledgement

The authors are grateful to Qiang Gao, Chen Shen and Ilias Samathrakis for useful discussion and suggestions. The authors gratefully acknowledge computational time on the Lichtenberg High Performance Supercomputer. Teng Long thanks the financial support from the China Scholarship Council. Nuno M. Fortunato thanks the financial support from European Research Council.

## Note

During the preparation of the manuscript, we noticed that Nelson and Sanvito did a similar work "Predicting the Curie temperature of ferromagnets using machine learning" (arXiv:1906:08534). They focused on the preselected FM compounds without classification, where the accuracy is about 88% with a MAE of 50K. Consistent with our observation, it is concluded that only chemistry is required to model the Curie temperature.